\documentclass[12pt,notitlepage]{iopart}

\usepackage{iopams}
\usepackage{graphicx}
\usepackage{braket}
\usepackage{float}
\usepackage{amsmath}
\usepackage{color}

\begin{document}

\title{Composite fermion-boson mapping for fermionic lattice models}
\author{J. Zhao$^1$, C. A. Jim\'enez-Hoyos$^1$, G. E. Scuseria$^{1,2}$, D. Huerga$^3$, J. Dukelsky$^3$, S. M. A. Rombouts$^4$, and G. Ortiz$^5$}

\address{$^1$ Department of Chemistry, Rice University, Houston, Texas 77005, USA}
\address{$^2$ Department of Physics and Astronomy, Rice University, Houston, Texas 77005, USA}
\address{$^3$ Instituto de Estructura de la Materia, CSIC, Serrano 123, E-28006 Madrid, Spain}
\address{$^4$ Departamento de F\'{i}sica Aplicada, Universidad de Huelva, E-21071 Huelva, Spain}
\address{$^5$ Department of Physics, Indiana University, Bloomington IN 47405, USA}

\begin{abstract}
We present a mapping of elementary fermion operators onto a quadratic form of
composite fermionic and bosonic cluster operators. The mapping is an exact
isomorphism as long as the physical constraint of one composite particle per
cluster is satisfied. This condition is treated on average in a composite
particle mean-field approach, which consists of an ansatz that decouples the
composite fermionic and bosonic sectors. The theory is tested on the one- and
two-dimensional Hubbard models.  Using a Bogoliubov determinant for the
composite fermions and either a coherent or Bogoliubov state for the bosons, we
obtain a simple and accurate procedure for treating the Mott insulating phase
of the Hubbard model with mean-field computational cost.
\end{abstract}

\section{Introduction}

The Hubbard model is a prototypical example of a strongly correlated system
characterized by the competition between the strong particle interaction ($U$)
and the kinetic energy ($t$). It is exactly solvable in one dimension, where
the ground state at half filling is a Mott insulator for any repulsive non-zero
interaction \cite{LiebAndWu}. Despite the simplicity of the model, the physics
arising in dimensions higher than one remains poorly understood. Different
many-body approximations have been applied along the years in different lattice
geometries and coupling regimes ($U/t$). Among them, let us cite Quantum and
Variational Monte Carlo calculations \cite{Guber, Guerre, Sorella}, Dynamical
Mean-Field Theory \cite{Roze}, Density Matrix Renormalization Group
\cite{Scholl} and, more recently, Density Matrix Embedding Theory \cite{Chan,
Irek}.  However, all these approaches have shown their limitations to describe
the strongly correlated regime of the Hubbard model ($U/t \gg 1$), in spite of
the significant computational cost of most of them. In order to overcome this
limitation, DMFT theories have been extended to clusters, showing excellent
convergence properties in one and two dimensional lattices \cite{Cellular,
Balzer}.  Alternative approaches, based on slave-particle methods, were
developed in condensed matter physics to model strongly correlated systems
\cite{Kotliar, Zou, Ostlund}. These methods, which treat a transformed
boson-fermion Hamiltonian, may provide access to a much better approximation to
the true ground-state and to physical processes that are otherwise difficult to
build into the many-body wave function \cite{Wen}.

Slave particle mappings may not be canonical; i.e., they may not preserve the
commutation properties of the mapped operators. Exact nonlinear mappings have
been studied sparingly \cite{Holten}. Even if the mapping is canonical, mixtures
with unphysical states may appear in the wave function due to the mean-field
approximations often employed to treat the slave-particle Hamiltonians. These
states need to be removed, albeit only approximately in practice. There are
numerous examples of slave-particle mappings in the literature.  In this
contribution, we explore a generalization to clusters of the Zou-Anderson (ZA)
mapping \cite{Zou, Ribeiro}. The many-body states of each cluster Fock space
will be mapped onto \emph{composite} boson and fermions operators. This
composite character has been already revealed in previous cluster mappings of
spin systems to composite bosons \cite{Ortiz}.  More recently, we have extended
this mapping to lattice boson systems and applied it to the Bose-Hubbard
Hamiltonian with good success \cite{Huerga}. As a precursor to the formalism
that we present here, we mention the boson-fermion plaquette model approach of
Altman and Auerbach \cite{Altman} for the 2D Hubbard model. There, a few  states
of bosonic and fermionic character were selected ad hoc and then treated within
the contractor renormalization group. On the contrary, the Composite
Fermion-Boson (CFB) mapping that we present in this work takes into account
{\textit{all}} states of the cluster, recovering the ZA mapping in the one-site
cluster limit.

\section{Theory}

\subsection{Fermion-boson composite mapping}

Let us start our derivation by decomposing the original lattice into a perfectly
tiled cluster lattice that will be referred to as the superlattice. Preferably,
the geometry of the clusters will be chosen such that they preserve as much as
possible the symmetries of the lattice. The Fock space of the complete system
$\mathbb{F}$ is a direct product of the Fock spaces of each cluster
$\mathbb{F}_\mathbf{R}$, where $\mathbf{R}$ denotes the position of the cluster
in the superlattice. The states in $\mathbb{F}_\mathbf{R}$  contain the complete
set of many-body states up to the maximum number of fermions that the cluster
can accommodate. Each cluster Fock space is decomposed into two subspaces with
odd and even number of fermions, which are denoted by $\mathbb{F}^o_\mathbf{R}$
and $\mathbb{F}^e_\mathbf{R}$. Each state of $\mathbb{F}^o_\mathbf{R}$ and
$\mathbb{F}^e_\mathbf{R}$ can be represented by the action of a
\textit{composite fermion } (CF)  and a \textit{composite boson } (CB)
operators, respectively, over the corresponding vacuum,
\begin{equation}
\vert \mathbf{R}\alpha\rangle\rightarrow a_{\mathbf{R}\alpha}^{\dag}\lvert 0\rangle,~~~~
\vert \mathbf{R}\beta\rangle\rightarrow b_{\mathbf{R}\beta}^{\dag}\lvert 0\rangle,
\end{equation}
where $\alpha~(\beta)$ labels clusters states with an odd (even) number of
fermions, respectively. The new composite particle Fock space is larger than the
original physical space. However, the subspace defined by all states having
one-and-only-one composite particle on each superlattice site has a one-to-one
correspondence with the original physical space. This is equivalent to require
\begin{equation}
\sum_{\alpha} a^{\dag}_{\mathbf{R}\alpha}a_{\mathbf{R}\alpha}+\sum_{\beta} b^{\dag}_{\mathbf{R}\beta}b_{\mathbf{R}\beta}=I, \label{physical_constraint}
\end{equation}
at each superlattice site $\mathbf{R}$. This condition will be referred to as
the \textit{physical constraint}. The formal mapping which relates the physical
fermionic operators with the new composite particles reads
\begin{equation}
c^{\dag}_{j\sigma}=\sum_{\alpha\beta}\langle \mathbf{R}\alpha\vert c^{\dag}_{j\sigma}\vert \mathbf{R}\beta\rangle ~a^{\dag}_{\mathbf{R}\alpha}b_{\mathbf{R}\beta} +
\sum_{\alpha\beta}\langle \mathbf{R}\beta\vert c^{\dag}_{j\sigma}\vert \mathbf{R}\alpha\rangle ~b^{\dag}_{\mathbf{R}\beta} a_{\mathbf{R}\alpha},~~~c_{j\sigma}=(c^{\dag}_{j\sigma})^{\dag}\label{CBF_map}
\end{equation}
where the site $j$ of the original lattice is contained within the cluster
$\mathbf{R}$ after the tiling. Notice that $\alpha$ and $\beta$ states in the
previous matrix elements differ by just one electron. The composite fermion
operators $( a^{\dag}_{\mathbf{R}\alpha},a_{\mathbf{R}\alpha} )$ satisfy the
anticommutation rules,
\begin{equation}
\{a_{\mathbf{R}\alpha}^\dag, a_{\mathbf{R}'\alpha^\prime}\} = \delta_{\alpha, \alpha^\prime }\delta_{\mathbf{R}\mathbf{R}'},
\qquad \{a_{\mathbf{R}\alpha}^\dag, a_{\mathbf{R}'\alpha^\prime}^\dag \}=0,\label{CF_com}
\end{equation}
while the composite boson operators $(
b^{\dag}_{\mathbf{R}\beta},b_{\mathbf{R}\beta} )$ satisfy bosonic commutation
rules
\begin{equation}
\lbrack b_{\mathbf{R}\beta}, b_{\mathbf{R}'\beta^\prime}^\dag \rbrack = \delta_{\beta \beta^\prime }\delta_{\mathbf{R}\mathbf{R}'},
\qquad \lbrack b_{\mathbf{R}\beta}^\dag, b_{\mathbf{R}'\beta^\prime}^\dag \rbrack = 0.\label{CB_com}
\end{equation}
The composite bosons and fermions commute with each other,
\begin{equation}
\lbrack a_{\mathbf{R}\alpha}, b_{\mathbf{R}\beta}^\dag \rbrack = 0.\label{CBF_com}
\end{equation}

Let us now explore the conditions that should be fulfilled by transformation
(\ref{CBF_map}) in order to preserve the canonical fermionic anticommutation
relations, $\lbrace
c_{i\sigma},c^{\dag}_{j\sigma'}\rbrace=\delta_{ij}\delta_{\sigma\sigma'}$. For
$i,j\in \mathbf{R}$, we insert the transformation (\ref{CBF_map}) into the
commutator and obtain,
\begin{eqnarray}
\lbrace c_{i\sigma},c^{\dag}_{j\sigma'}\rbrace &=&
\sum_{\alpha\alpha'}\sum_{\beta'} \langle \mathbf{R}\alpha\vert c_{i\sigma}\vert \mathbf{R}\beta'\rangle
\langle \mathbf{R}\beta'\vert c^{\dag}_{j\sigma'}\vert \mathbf{R}\alpha'\rangle
~a^{\dag}_{\mathbf{R}\alpha} a_{\mathbf{R}\alpha'}\notag\\
&&+\sum_{\beta\beta'}\sum_{\alpha'} \langle \mathbf{R}\beta\vert c_{i\sigma}\vert \mathbf{R}\alpha'\rangle
\langle \mathbf{R}\alpha'\vert c^{\dag}_{j\sigma'}\vert \mathbf{R}\beta'\rangle
~b^{\dag}_{\mathbf{R}\beta} b_{\mathbf{R}\beta'}\notag\\
&&+\sum_{\alpha\alpha'}\sum_{\beta'} \langle \mathbf{R}\alpha\vert c_{j\sigma'}^{\dag}\vert \mathbf{R}\beta'\rangle
\langle \mathbf{R}\beta'\vert c_{i\sigma}\vert \mathbf{R}\alpha'\rangle
~a^{\dag}_{\mathbf{R}\alpha} a_{\mathbf{R}\alpha'}\notag\\
&&+\sum_{\beta\beta'}\sum_{\alpha'} \langle \mathbf{R}\beta\vert c_{j\sigma'}^{\dag}\vert \mathbf{R}\alpha'\rangle
\langle \mathbf{R}\alpha'\vert c_{i\sigma}\vert \mathbf{R}\beta'\rangle
~b^{\dag}_{\mathbf{R}\beta} b_{\mathbf{R}\beta'}.\label{cluster_com}
\end{eqnarray}
where we have used the commutation relations (\ref{CF_com}), (\ref{CB_com}), and
applied the physical constraint (\ref{physical_constraint}). Noting that the
complete set of bosonic and fermionic cluster states satisfy a resolution of the
identity,
\begin{equation}
\sum_{\beta}\vert \mathbf{R}\beta\rangle \langle \mathbf{R}\beta\vert + \sum_{\alpha}\vert \mathbf{R}\alpha\rangle \langle \mathbf{R}\alpha\vert=I,\label{resol_identity}
\end{equation}
and taking into account that the matrix elements $\langle \mathbf{R}\alpha\vert
c^{\dag}_{j\sigma} \vert \mathbf{R}\alpha'\rangle$, $\langle
\mathbf{R}\alpha\vert c_{j\sigma} \vert \mathbf{R}\alpha'\rangle$, $\langle
\mathbf{R}\alpha\vert c^\dag_{j\sigma} \vert \mathbf{R}\alpha'\rangle$ and
$\langle \mathbf{R}\beta\vert c_{j\sigma} \vert \mathbf{R}\beta'\rangle$ vanish,
equation (\ref{cluster_com}) reduces to
\begin{eqnarray}
\lbrace c_{i\sigma}, c_{j\sigma'}^{\dag}\rbrace&=&
\sum_{\alpha\alpha'}\langle \mathbf{R}\alpha \vert \lbrace c_{i\sigma}, c_{j\sigma'}^{\dag}\rbrace \vert \mathbf{R}\alpha'\rangle a_{\mathbf{R}\alpha}^{\dag} a_{\mathbf{R}\alpha'}\notag\\
&&+\sum_{\beta\beta'}\langle \mathbf{R}\beta \vert \lbrace c_{i\sigma}, c_{j\sigma'}^{\dag}\rbrace \vert \mathbf{R}\beta'\rangle b_{\mathbf{R}\beta}^{\dag} b_{\mathbf{R}\beta'}\\
&=& \delta_{i,j} \delta_{\sigma, \sigma'},
\end{eqnarray}
where in the last step we have made use of the anticommutation relations of the
physical fermions, the orthogonality of the cluster basis, and the physical
constraint (\ref{physical_constraint}). Note that the result is equivalent to a
direct mapping of $\lbrace c_{i\sigma}, c_{j\sigma'}^{\dag} \rbrace$ onto the
composite particle space. The equivalence of both mappings is maintained in the
physical space, characterized by the satisfaction of the physical constraint
(\ref{physical_constraint}), when the complete Fock space is used in each
cluster. We also point out that the anticommutation relation $\lbrace
c_{i\sigma},c^{\dag}_{j\sigma'}\rbrace=0$ for $i\in \mathbf{R}, j\in
\mathbf{R}'\neq \mathbf{R}$ is trivially satisfied by using the commutation
relations (\ref{CF_com}), (\ref{CB_com}) and (\ref{CBF_com}).

\begin{figure}
\begin{center}
\includegraphics[scale=0.3]{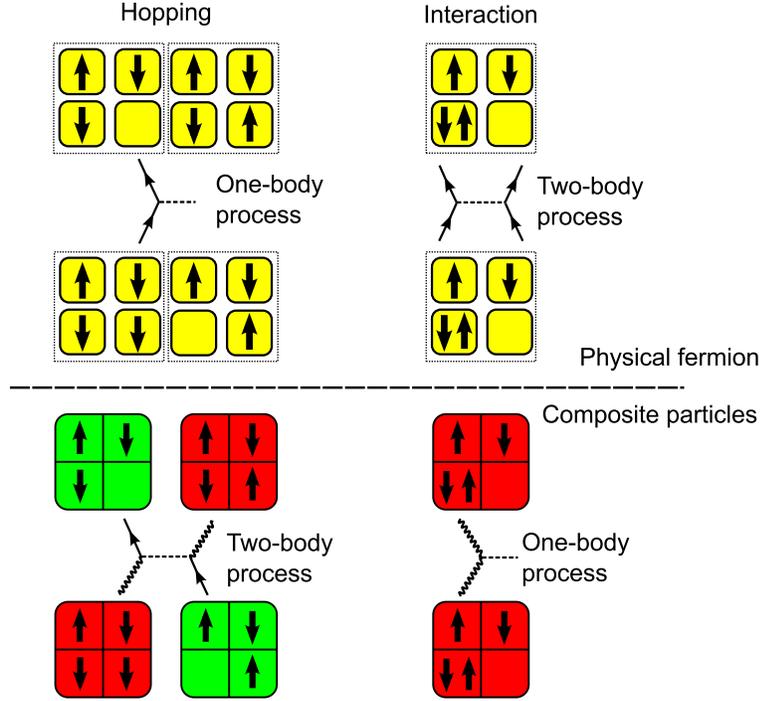}
\end{center}
\caption{Schematic representation of an inter-cluster hopping and an on-site
interaction processes before (upper part of the figure) and after a mapping to
$2 \times 2$ composite particles (lower part of the figures).  The small yellow
squares represent the original lattice sites. Green $2 \times 2$ clusters
represents CF states, while the red clusters represent CB states.  The figure
illustrates how a inter-cluster hopping process maps onto a two-body CFB
interaction while an on-site two-body interaction maps onto a one-body CB or CF
term.}
\label{fig:br}
\end{figure}

As long as the complete set of bosonic and fermionic cluster states is used,
one can map an arbitrary operator acting within a cluster $\mathbf{R}$ to a
one-body composite operator of the general form
\begin{eqnarray} \hat{\mathcal{O}}_\mathbf{R}&=& \sum_{\alpha\alpha'}\langle
\mathbf{R}\alpha\vert \hat{\mathcal{O}}_\mathbf{R} \vert \mathbf{R}\alpha'
\rangle a^{\dag}_{\mathbf{R}\alpha}a_{\mathbf{R}\alpha'} +
\sum_{\beta\beta'}\langle \mathbf{R}\beta\vert \hat{\mathcal{O}}_\mathbf{R}
\vert \mathbf{R}\beta' \rangle b^{\dag}_{\mathbf{R}\beta}b_{\mathbf{R}\beta'}
\notag\\ &&+\sum_{\alpha\beta}\langle \mathbf{R}\beta\vert
\hat{\mathcal{O}}_\mathbf{R} \vert \mathbf{R}\alpha \rangle
b^{\dag}_{\mathbf{R}\beta}a_{\mathbf{R}\alpha} +\sum_{\alpha\beta}\langle
\mathbf{R}\alpha\vert \hat{\mathcal{O}}_\mathbf{R} \vert \mathbf{R}\beta
\rangle a^{\dag}_{\mathbf{R}\alpha}b_{\mathbf{R}\beta}.  \end{eqnarray}
The first line applies if the operator $\hat{\mathcal{O}}_\mathbf{R}$ preserves
the number of fermions, or creates or annihilates an even number of fermions
(even number parity). The second line applies if it creates or annihilates an
odd number of fermions (odd number parity).  Equivalently, any algebraic
operator acting on $n$ different clusters will be mapped to a general $n$-body
CFB operator. As an example, let us apply the mapping (\ref{CBF_map}) to the
Hubbard Hamiltonian in a hypercubic lattice with $N$ sites in $d$ dimensions,
\begin{equation} \hat{H} = -t \sum_{\langle i,j \rangle,\sigma}(
c^{\dagger}_{i,\sigma} c^{}_{j,\sigma}+ c^{\dagger}_{j,\sigma}c^{}_{i,\sigma})
+ U\sum_{i=1}^{N}  n_{i\uparrow}  n_{i\downarrow}
 - \mu \sum_{i=1}^N n_i.\label{Hub_Ham} \end{equation}
where $c_{i\sigma}^{\dag}~(c_{i\sigma})$ creates (annihilates) a fermion at
lattice site $i$ with spin $\sigma = \uparrow, \downarrow$, and $n_{i\sigma}=
c_{i\sigma}^{\dag}c_{i\sigma}$ is the number operator,
$n_{i}=n_{i\uparrow}+n_{i\downarrow}$. The first term accounts for the hopping
of fermions to nearest neighbor sites with tunneling amplitude $t$. The second
term accounts for the on-site interaction of strength $U$, and the third term
regulates the density of the system via an external chemical potential $\mu$.
In what follows, we express all quantities in units of the hopping parameter
$t$.  In terms of the composite particles, the mapped Hamiltonian is
\begin{eqnarray} \hat{H}_{CFB}&=& \sum_{\mathbf{R}}
\left[\left(T_{\mathbf{R}}\right)^{\alpha}_{\alpha'}
a^{\dag}_{\mathbf{R}\alpha} a_{\mathbf{R}\alpha'}
+\left(T_{\mathbf{R}}\right)^{\beta}_{\beta'} b^{\dag}_{\mathbf{R}\beta}
b_{\mathbf{R}\beta'}\right]\notag\\ &&+\sum_{\langle
\mathbf{R}\mathbf{R}'\rangle}\left[\left(V_{\mathbf{R}\mathbf{R}'}\right)^{\alpha\beta}_{\beta'\alpha'}
a^{\dag}_{\mathbf{R}\alpha}b^{\dag}_{\mathbf{R}'\beta}b_{\mathbf{R}\beta'}a_{\mathbf{R}'\alpha'}
+ \text{H.c.}\right]\notag\\ &&+\sum_{\langle
\mathbf{R}\mathbf{R}'\rangle}\left[\left(V_{\mathbf{R}\mathbf{R}'}\right)^{\alpha\alpha'}_{\beta\beta'}
a^{\dag}_{\mathbf{R}\alpha}a^{\dag}_{\mathbf{R}'\alpha'}b_{\mathbf{R}\beta}b_{\mathbf{R}'\beta'}
+\text{H.c.}\right]\label{CBF_Ham} \end{eqnarray}
where repeated Greek indices sum. The tensors $T$ and $V$, defined as
\begin{eqnarray} \left(T_\mathbf{R}\right)^{\alpha}_{\alpha'}&=& \langle
\mathbf{R}\alpha\vert \hat{H}^{\square}_\mathbf{R}\vert
\mathbf{R}\alpha'\rangle\\ \left(T_\mathbf{R}\right)^{\beta}_{\beta'}&=&
\langle \mathbf{R}\beta\vert \hat{H}^{\square}_\mathbf{R}\vert
\mathbf{R}\beta'\rangle\\
\left(V_{\mathbf{R}\mathbf{R}'}\right)^{\alpha\beta}_{\beta'\alpha'}&=& \langle
\mathbf{R}\alpha,\mathbf{R}'\beta\vert \hat{H}_{\mathbf{R}\mathbf{R}'}^{\times}
\vert \mathbf{R}\beta',\mathbf{R}'\alpha'\rangle\\
\left(V_{\mathbf{R}\mathbf{R}'}\right)^{\alpha\alpha'}_{\beta\beta'}&=& \langle
\mathbf{R}\alpha,\mathbf{R}'\alpha'\vert
\hat{H}_{\mathbf{R}\mathbf{R}'}^{\times} \vert
\mathbf{R}\beta,\mathbf{R}'\beta'\rangle \end{eqnarray}
contain all the information about the original Hamiltonian (\ref{Hub_Ham}).
Here, $\hat{H}_{\mathbf{R}}^{\square}$ is the part of the Hubbard Hamiltonian
acting within a cluster $\mathbf{R}$, and the interaction between neighbor
clusters $\hat{H}_{\mathbf{R}\mathbf{R}'}^{\times}$ is exclusively due to the
hopping term.  Notice that due to the hermiticity of the interaction $V$, the
tensors are symmetric under certain interchange of its indices,
\begin{eqnarray} (V_{\mathbf{R}\mathbf{R}'})^{\beta\alpha}_{\alpha'\beta'}&=&
\bra{\mathbf{R}\beta,\mathbf{R}'\alpha}
\hat{H}^{\times}_{\mathbf{R}\mathbf{R}'}\ket{\mathbf{R}\alpha',\mathbf{R}'\beta'}=
(V^\ast_{\mathbf{R}\mathbf{R}'})_{\beta\alpha}^{\alpha'\beta'}~,\\
(V_{\mathbf{R}\mathbf{R}'})^{\alpha\alpha'}_{\beta\beta'}&=&
\bra{\mathbf{R}\alpha,\mathbf{R}'\alpha'}
\hat{H}^{\times}_{\mathbf{R}\mathbf{R}'}\ket{\mathbf{R}\beta,\mathbf{R}'\beta'}=
(V^\ast_{\mathbf{R}\mathbf{R}'})_{\alpha\alpha'}^{\beta\beta'}~.  \end{eqnarray}
In the following, as we have only considered uniform, periodic superlattices
with only nearest neighbour hopping, we will drop the $\mathbf{R}$ labels on
the tensor $T$ since it has the same value for all the clusters. In the same
vein, the $V$ tensor vanishes for all non-neighbour cluster pairs and takes the
same value for all cluster pairs neighbouring in the same orientation. As a
result, we shall drop the full $\mathbf{R} \mathbf{R}'$ subscript and just
subscript it with the neighbouring orientation. For instance, $V_x$ denotes the
$V$ tensor for clusters neighbouring in the $x$ direction.  Similarly to the ZA
mapping, which is the one site cluster limit, the body-rank of hopping and
on-site interaction terms are swapped. The one-body inter-cluster hopping is
mapped into a two-body CFB term involving fermion-boson interactions, while the
original two-body on-site interaction is mapped into a sum of one-body
composite fermion and one-body composite boson terms. As an illustration, this
swapping effect is schematically shown in Figure \ref{fig:br} through an
example of an inter-cluster hopping process and an on-site interaction on
$2\times2$ neighbor clusters.

The Hamiltonian (\ref{CBF_Ham}) is an exact isomorphism of the Hubbard Hamiltonian
(\ref{Hub_Ham}) within the physical subspace of the composite particle Fock
space. Although the complexity of the original problem is not reduced, suitable
approximations can be carried out with the advantage that short-range quantum
correlations are included automatically within the collective structure of the
composite particles.

\subsection{Mean-field solution}

In order to proceed further, we assume that the system is translationally
invariant and thus we perform a discrete Fourier transform of the composite
operators,
\begin{eqnarray} a_{\mathbf{R}\alpha }^{\dag }
&=&\frac{1}{\sqrt{M}}\sum_{\mathbf{K}\in BZ}e^{-{\rm
i}\mathbf{K}\mathbf{R}}a_{\mathbf{K}\alpha}^{\dag },\\ b_{\mathbf{R}\beta
}^{\dag } &=&\frac{1}{\sqrt{M}}\sum_{\mathbf{K}\in BZ}e^{-{\rm
i}\mathbf{K}\mathbf{R}}b_{\mathbf{K}\beta }^{\dag}, \end{eqnarray}
where $M$ is the number of sites of the hypercubic superlattice of clusters
with size $L^d$. The first Brillouin zone ($BZ$) is defined in the interval
$(-\pi/L,\pi/L]$ in each direction of the hypercubic reciprocal space. Under
this transformation, the Hamiltonian can be written as
\begin{eqnarray} \hat{H}_{CFB} &=& \sum_{\mathbf{K}}\left( T_{\beta'}^{\beta}
b^{\dag}_{\mathbf{K}\beta}b_{\mathbf{K}\beta'}+ T_{\alpha'}^{\alpha}
a^{\dag}_{\mathbf{K}\alpha} a_{\mathbf{K}\alpha'}\right) \notag\\
&&+\frac{1}{M}\sum_u  \sum_{ \mathbf{K_{1}K_{2}Q} }
\left(V_u\right)_{\alpha'\beta'}^{\beta\alpha}\notag\\ &&~~~~\times\left(
e^{-{\rm i}Q_{u}L} b_{\mathbf{K_1},\beta}^{\dag} a_{\mathbf{K_2+Q},\alpha
}^{\dag} a_{\mathbf{K_1+Q},\alpha'} b_{\mathbf{K_2},\beta'} + \text{H.c.}
\right)\notag\\ &&+\frac{1}{M}\sum_u \sum_{ \mathbf{K_{1}K_{2} Q} }
\left(V_u\right)_{\beta\beta'}^{\alpha\alpha'}\notag\\ &&~~~~\times\left(
e^{-{\rm i}Q_{u}L} a_{\mathbf{K_1},\alpha}^{\dag}
a_{\mathbf{K_2+Q},\alpha'}^{\dag} b_{\mathbf{K_1+Q},\beta}
b_{\mathbf{K_2},\beta'} + \text{H.c.} \right)\label{HK} \end{eqnarray}
where repeated Greek indices sum and $\mathbf{K}_1$, $\mathbf{K}_2$,
$\mathbf{K}$, and $\mathbf{Q}$ are summed over all the vectors in the Brillouin
zone. $u$ is summed over the spacial dimensions of the system and $V_u$ denotes
the $V$ tensor for cluster pairs neighbouring in the direction of $u$.

The mean-field treatment of the Hamiltonian (\ref{HK}) assumes a decoupling of
the composite fermion and boson spaces, $\ket{\Psi} = \ket{\Psi^{F}} \otimes
\ket{\Psi^{B}}$, yielding an effective mean-field Hamiltonian that is quadratic
in both the fermionic and bosonic sectors and can be diagonalized by two
independent fermionic and bosonic Bogoliubov transformations. The two sectors
are coupled via self-consistent boson and fermion mean-fields. In other words,
by taking partial variations of the expectation value of $H_{CFB}$ with respect
to the fermion and boson wave functions, the stationary condition implies that
$\ket{\Psi^{F}}$ and $\ket{\Psi^{B}}$ are eigenfunctions of the effective
fermion and boson Hamiltonians,
\begin{eqnarray}
\hat{H}^{F} & = & \bra{\Psi^{B}} \hat{H}_{CFB} \ket{\Psi^{B}},\\
\hat{H}^{B} & = & \bra{\Psi^{F}} \hat{H}_{CFB} \ket{\Psi^{F}}.
\label{eqn:mf}
\end{eqnarray}
More precisely, the fermionic sector reads
\begin{align}
\hat{H}^{F}&=
\sum_{\mathbf{K}}
T_{\alpha'}^{\alpha} a^{\dag}_{\mathbf{K}\alpha} a_{\mathbf{K}\alpha'}\notag\\
&+\frac{1}{M}  \sum_u\sum_{ \mathbf{KQ} } 
\left(V_u\right)_{\alpha'\beta'}^{\beta\alpha}
\left(
e^{-{\rm i}(K_u-Q_u)}
a_{\mathbf{K},\alpha }^{\dag}
a_{\mathbf{K},\alpha'}
\langle b_{\mathbf{Q},\beta}^{\dag}
b_{\mathbf{Q},\beta'}\rangle + \text{H.c.}
\right)\notag\\
&+\frac{1}{M}  \sum_u \sum_{ \mathbf{KQ} } 
\left(V_u\right)_{\beta\beta'}^{\alpha\alpha'}
\left(
e^{-{\rm i}(K_u-Q_u)}
a_{\mathbf{K},\alpha}^{\dag}
a_{\mathbf{-K},\alpha'}^{\dag}
\langle b_{\mathbf{Q},\beta}
b_{\mathbf{-Q},\beta'}\rangle + \text{H.c.}
\right),
\label{HCBF_fermion_mf}
\end{align}
and the bosonic sector is obtained by a counterpart mean-field decoupling,
\begin{align}
\hat{H}^{B}&=
\sum_{\mathbf{K}}
T_{\beta'}^{\beta} b^{\dag}_{\mathbf{K}\beta} b_{\mathbf{K}\beta'}\notag\\
&+\frac{1}{M}  \sum_u\sum_{ \mathbf{KQ} } 
\left(V_u\right)_{\alpha'\beta'}^{\beta\alpha}
\left(
e^{-{\rm i}(K_u-Q_u)}
b_{\mathbf{K},\beta}^{\dag}
b_{\mathbf{K},\beta'}
\langle a_{\mathbf{Q},\alpha }^{\dag}
a_{\mathbf{Q},\alpha'}\rangle
+ \text{H.c.}
\right)\notag\\
&+\frac{1}{M}  \sum_u\sum_{ \mathbf{KQ} } 
\left(V_u\right)_{\beta\beta'}^{\alpha\alpha'}
\left(
e^{-{\rm i}(K_u-Q_u)}
b_{\mathbf{K},\beta}
b_{\mathbf{-K},\beta'}
\langle
a_{\mathbf{Q},\alpha}^{\dag}
a_{\mathbf{-Q},\alpha'}^{\dag}\rangle
 + \text{H.c.}
\right).
\label{HCBF_boson_mf}
\end{align}
The physical constraint implies that one-and-only-one state is allowed per
cluster, however, the mean-field decoupling of Eqs. (\ref{HCBF_fermion_mf}) and
(\ref{HCBF_boson_mf}) leads to wave functions $\ket{\Psi^{F}}$ and
$\ket{\Psi^{B}}$ which do not preserve the local physical constraint
(\ref{physical_constraint}) exactly.  Therefore, we relax it fixing a global
constraint of the total composite particle density,
\begin{equation}
\frac{1}{M}\sum_{\mathbf{R}}\left(
\sum_{\alpha} \braket{ a^{\dag}_{\mathbf{R}\alpha} a_{\mathbf{R}\alpha} }
+
\sum_{\beta} \braket{ b^{\dag}_{\mathbf{R}\beta} b_{\mathbf{R}\beta} }
\right)=1.
\end{equation}
This latter condition is added to the effective bosonic and fermionic
Hamiltonians via a unique Lagrange multiplier $\lambda$,
\begin{eqnarray}
\hat{F}^{F}&=& \hat{H}^{F} -\lambda \sum_{\mathbf{K}} \sum_{\alpha}
a^{\dag}_{\mathbf{K}\alpha} a_{\mathbf{K}\alpha},\label{F_F}\\
\hat{F}^{B}&=& \hat{H}^{B} -\lambda \sum_{\mathbf{K}} \sum_{\beta}
b^{\dag}_{\mathbf{K}\beta} b_{\mathbf{K}\beta}.\label{F_B}
\end{eqnarray}
In this work, we have restricted ourselves to wave functions transforming as the
totally-symmetric representation of the lattice translation group of the cluster
superlattice. We would like to point out that, when the wave function does not
break the translational symmetry of the cluster superlattice, the global
imposition of the physical constraint would imply local on-average satisfaction
of the physical constraint,
\begin{equation}
\sum_{\alpha} \braket{ a^{\dag}_{\mathbf{R}\alpha} a_{\mathbf{R}\alpha} }
+
\sum_{\beta} \braket{ b^{\dag}_{\mathbf{R}\beta} b_{\mathbf{R}\beta} }
=1,
\label{physical-average}
\end{equation}
for all clusters $\mathbf{R}$. Like in other slave particle approaches,
fluctuations of the physical constraint induce mixtures with unphysical states.
These unphysical processes are expected to decrease with increasing cluster
sizes, approaching the exact physical eigenstate in the infinite size limit, in
spite of the combinatorial increase of the computational cost.

In the same vein as our previous composite particle treatment of boson systems
\cite{Huerga}, one can consider several candidate mean-field reference states.
The fermionic part of the product wave function may be treated within a
Hartree-Fock (HF) or a Hartree-Fock-Bogoliubov (HFB) approximation, while the
bosonic part may be treated within a Hartree-Bose (HB) or a
Hartree-Bose-Bogoliubov (HBB) approximation.  In this work, we present results
for the Hubbard model in one and two dimensions obtained by diagonalizing the
fermionic sector via a general HFB transformation, and by treating the bosonic
sector in HB and HBB approximations. In the first approximation, we assume that
the bosonic sector is described by a coherent state of CBs in the
$\mathbf{K=0},~\alpha=\sf{c}$ mode, and neglect all bosonic fluctuations (HB).
In the second approximation, the bosonic sector is also diagonalized by means of
a Bogoliubov transformation (HBB).

Within the coherent approximation, we replace the condensate of CB by a
$c-$number $b^{(\dag)}_{\mathbf{0},\sf{c}}\rightarrow \sqrt{M}\sigma$, where
$\sigma^2$ is the CB \textit{condensate fraction}. The coherent CB is a linear
combination of CB configurations, i.e., $\ket{\sf{c}}=\sum_{\beta}
U^{\sf{c}}_{\beta}\ket{\beta}$. Inserting this transformation into Eq.
(\ref{F_B}) and minimizing with respect to the variational amplitudes
$U^{\sf{c}}$ leads to the Hartree-Bose eigensystem, where $\lambda$ is the
lowest eigenvalue, and $U^{\sf{c}}$ its corresponding eigenvector,
\begin{equation}
\sum_{\beta'}h^{\mathbf{0},\mathbf{0}}_{\beta\beta'} U_{\beta'}^{\sf{c}}= \lambda U^{\sf{c}}_{\beta},
\end{equation}
where the superindexes of the Hartree matrix $h$ refer to superlattice momentum
$\mathbf{K}=\mathbf{0}$.
The value of the condensate fraction $\sigma^{2}$ is obtained through the
physical constraint,
\begin{equation}
\sigma^{2}=1 - \frac{1}{M}\sum_{\mathbf{K}\alpha}\langle a^{\dag}_{\mathbf{K}\alpha}a_{\mathbf{K}\alpha}\rangle.
\end{equation}

A general symmetry-preserving Bogoliubov transformation for quadratic
Hamiltonians has the form \cite{Ripka},
\begin{equation} \gamma_{\mathbf{K}\eta}= \sum_{\eta'}
\left(X^{\mathbf{K}}_{\eta'\eta}\right)^\ast \;  p^{\dag}_{\mathbf{K}\eta'} +
\varepsilon \sum_{\eta'} \left(Y_{\eta'\eta}^{- \mathbf{K}}\right)^* \;
p_{\mathbf{-K}\eta'} , \end{equation}
where the operators $(p^{\dag}_{\mathbf{K}\eta},p_{\mathbf{K}\eta})$ refer to
fermions $(\varepsilon=+1)$ or bosons $(\varepsilon=-1)$. Accordingly, the
$\eta$ labels either CF states $\alpha$ or CB states $\beta$. The amplitudes
$X$ and $Y$ are obtained by solving the self-consistent matrix eigensystem of
the form \cite{Ripka},
\begin{equation} \left( \begin{array}{cc} h^{\mathbf{K},\mathbf{K}} &
\Delta^{\mathbf{K},\mathbf{-K}}\\ \varepsilon (\Delta^{\mathbf{-K},
\mathbf{K}})^{*} & \varepsilon (h^{\mathbf{-K},\mathbf{-K}})^{*}\end{array}
\right) \left( \begin{array}{cc} (X^{\mathbf{K}})^{*} & Y^{\mathbf{K}}\\
(Y^{- \mathbf{K}})^{*} & X^{- \mathbf{K}} \end{array} \right)=\Omega^{\mathbf{K}}
\left( \begin{array}{cc} (X^{\mathbf{K}})^{*} & Y^{\mathbf{K}}\\
(Y^{- \mathbf{K}})^{*} & X^{- \mathbf{K}} \end{array} \right)\label{Bogo_diag}
\end{equation}
where the positive eigenvalues of the diagonal matrix
$\Omega^{\mathbf{K}}=\text{Diag}(- w^{\mathbf{K}},w^{\mathbf{K}})$ will
determine the fermionic (bosonic) quasi-particle excitation dispersions.  The
matrix elements are straightforwardly obtained by identifying the
grand-canonical composite particle potentials given in (\ref{F_F}) and
(\ref{F_B}) with the general expression
\begin{equation} \hat{F}= \sum_{\mathbf{K}\eta\eta'}
\left(h_{\eta\eta'}^{\mathbf{K},\mathbf{K}}
p_{\mathbf{K}\eta}^{\dag}p_{\mathbf{K}\eta'}
\right)
+ \frac{1}{2} \sum_{\mathbf{K}\eta\eta'} \left( \Delta_{\eta\eta'}^{\mathbf{K},\mathbf{-K}}
p^{\dag}_{\mathbf{K}\eta}p^{\dag}_{\mathbf{-K}\eta'} + \text{H.c.}\right)
\end{equation}
The block diagonal structure with respect to $\mathbf{K}$ stems from the fact
that all normal density matrix elements $\braket{p^\dag_{\mathbf{K}_1\eta}
p_{\mathbf{K}_2 \eta}}$ vanish except for $\mathbf{K}_1 = \mathbf{K}_2$ and the
anomalous density matrix elements $\braket{p_{\mathbf{K}_1\eta}
p_{\mathbf{K}_2\eta}}$ vanish except for opposite momentum $\mathbf{K}_1 =
-\mathbf{K}_2$ when the wave function transforms according to an irreducible
representation of the superlattice translation group. Upon inversion of the
Bogoliubov transformation, we obtain the normal and pairing tensors
\begin{eqnarray} \langle p^{\dag}_{\mathbf{K}\eta}p_{\mathbf{K}\eta'}\rangle&=&
\sum_{\zeta}(X^{\mathbf{K}}_{\eta' \zeta})^{*} X^{\mathbf{K}}_{\zeta\eta},\\
\langle
p^{\dag}_{\mathbf{K}\eta}p^{\dag}_{-\mathbf{K}\eta'}\rangle&=& -\varepsilon \sum_{\zeta}  (Y^{- \mathbf{K}}_{\eta'\zeta})^\ast X^{\mathbf{K}}_{\zeta\eta}
.  \end{eqnarray}

The convergence of the procedure is determined by the self-consistency between
the fermion and boson Hamiltonians and their respective mean-field solutions.
In the scheme where we diagonalize both sectors by coupling the fermionic and
bosonic Bogoliubov eigensystems self-consistently, the value of the Lagrange
multiplier $\lambda$ is varied smoothly until the density of composite particles
equals one.

In both of the approximations here considered, the energy of the system can be
readily computed taking the expectation value of the CFB Hamiltonian (\ref{HK})
with our ansatz $ \ket{\Psi}=\ket{\Psi^{B}}\otimes\ket{\Psi^{F}}$. In
particular, for the coherent approximation the total energy reads
\begin{eqnarray} E&=& MT_{\sf c}^{\sf c} \sigma^{2}+
\sum_{\mathbf{K}}T_{\alpha'}^{\alpha} \langle a^{\dag}_{\mathbf{K}\alpha}
a_{\mathbf{K}\alpha'}\rangle \notag\\ &&+\sigma^{2}\sum_{ \mathbf{K} }
\sum_u \left(V_u\right) _{\alpha'{\sf c}}^{{\sf c}\alpha} \left(e^{-{\rm i}K_u} \langle
a_{\mathbf{K}\alpha }^{\dag}
a_{\mathbf{K}\alpha'}\rangle+\text{H.c.}\right)\notag\\ &&+\sigma^{2}\sum_{
\mathbf{K} } \sum_u \left(V_u\right)_{\sf cc}^{\alpha\alpha'} \left( e^{-{\rm i}K_u}
\langle a_{\mathbf{K}\alpha}^{\dag} a_{\mathbf{-K}\alpha'}^{\dag}\rangle +
\text{H.c.} \right). 
\label{GSE}
 \end{eqnarray}
where repeated indices are summed and we have performed a contraction of the
original tensors $T$ and $V$ with the boson condensate $\sf{c}$.  Notice that,
as the mean-field treatment does not preserve the physical constraint exactly,
the energy reached is not an upper bound to the exact ground state energy.
Nevertheless, by performing a finite-size scaling analysis one can give a
quantitative estimate of the exact result.

We are also interested in the double-occupation parameter, which gives a measure
of on-site physical fermion correlations and provides a qualitative indicator of
the Mott insulator transition. It is here computed in the CFB mean-field
approximations as
\begin{equation}
D_j=\langle n_{j\downarrow}n_{j\uparrow}\rangle
=
\sum_{\beta} \bra{\mathbf{R}\beta}n_{j\uparrow}n_{j\downarrow}\ket{\mathbf{R}\beta}\langle
b^{\dag}_{\mathbf{R}\beta}b_{\mathbf{R}\beta}\rangle
+ \sum_{\alpha} \bra{\mathbf{R}\alpha}n_{j\uparrow}n_{j\downarrow}\ket{\mathbf{R}\alpha}\langle
a^{\dag}_{\mathbf{R}\alpha}a_{\mathbf{R}\alpha}\rangle
,
\label{double}
\end{equation}
with $j\in\mathbf{R}$.

As a short recapitulation, we have presented a canonical mapping which allows
us to express a many-fermion lattice Hamiltonian as a many fermion-boson system
on a superlattice of clusters. The composite boson and fermion cluster
operators contain the exact quantum correlations inside the clusters. In order
for the mapping to be an exact isomorphism, a physical constraint has to be
satisfied. We proposed a mean-field ansatz for our mapped Hamiltonian that is a
direct product of fermionic and bosonic states that preserve the superlattice
translational symmetry. In the following, we present results obtained with this
approach.

\section{Results and discussion}

\begin{figure}[htbp]
\begin{center}
\includegraphics[width=0.49\textwidth, height=6cm]{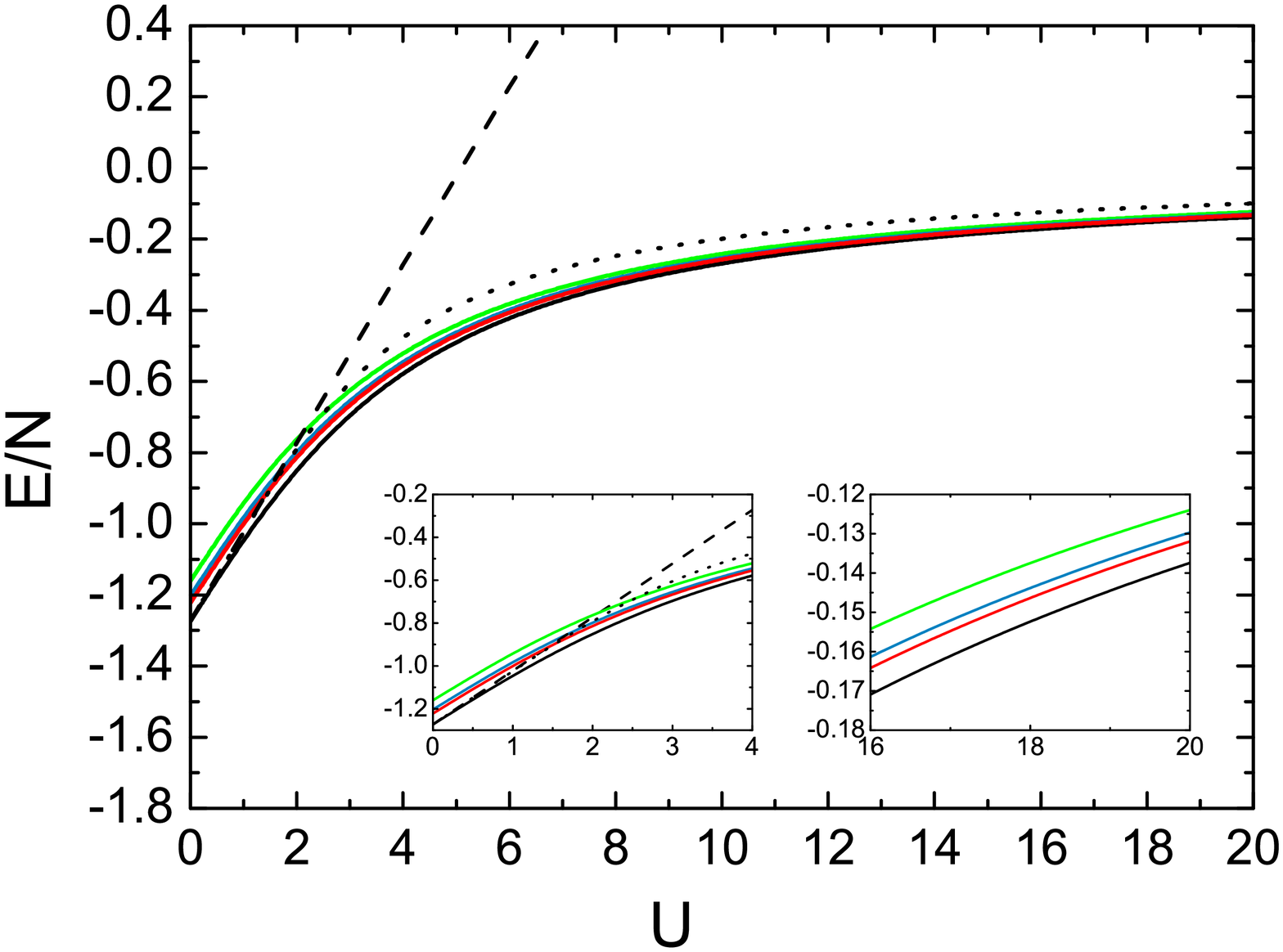}
\includegraphics[width=0.49\textwidth, height=6cm]{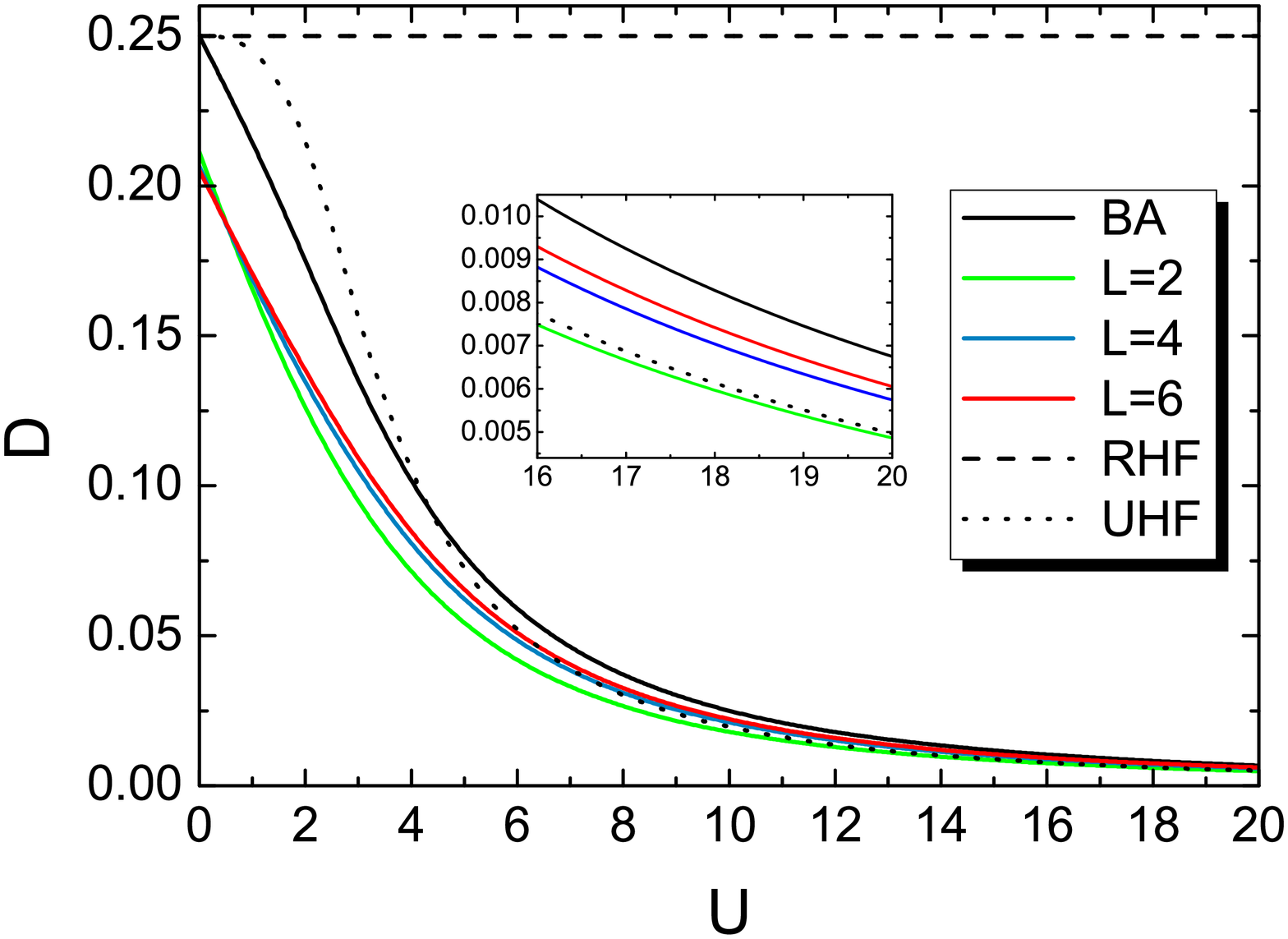}
\caption{ 
Ground state energies per site and double-occupancies of the 1-D Hubbard model
obtained from the CFB mean-field theory with various clusters sizes ($L=2,4,6$),
the restricted and unrestricted Hartree-Fock (HF) approximation, and the exact
Bethe anstaz (BA). The insets amplifies the differences in particular coupling
regimes $U$. 
}
\end{center}
\label{fig:oneDEU}
\end{figure}

We start the numerical analysis of the method by studying the ground-state
energy $E$ (\ref{GSE}) and double-occupation $D$ (\ref{double}) of the Hubbard
model at half-filling (\ref{Hub_Ham}) by means of the HB approximation for the
CB Hamiltonian (\ref{F_B}), and HFB for the CF Hamiltonian (\ref{F_F}).
Half-filling can be achieved by setting the chemical potential $\mu$ to half
the value of the Hubbard repulsion parameter $U$ for particle-hole symmetry. In
Figure \ref{fig:oneDEU}, we present the energy per site (left panel) and
double-occupation (right panel) for various values of $U$. We also include the
exact energy per site obtained by the Bethe ansatz \cite{LiebAndWu} and the
energy obtained by both the standard, symmetry-preserving  and the unrestricted
HF approximations of the original Hubbard Hamiltonian (\ref{Hub_Ham}).  The RHF
mean-field methods is exact in the non-interacting limit, $U=0$; however, it
quickly deteriorates with increasing interaction. The ground state energy
obtained with the CFB mean-field is in very good agreement with the exact
result for large $U$. The UHF method is able to provide a qualitatively
reasonable result for both small and large $U$. So it parallels the trend for
the Bethe ansatz, but becomes higher in energy than the CFB mean-field energy
for larger $U$ values. The right inset shows how the CFB mean-field energy
converges to the exact one as the size of the cluster is increased. In this
way, a significant portion of the correlation energy relative to the  RHF
solution can be captured. However, at low $U$, the energy from the CFB
mean-field theory starts to deviate from the exact Bethe ansatz value, and
crosses the  RHF energy at $U\sim 1.4$ for the size-six cluster, as it can be
seen in the left inset.

By inspecting the double-occupation expectation value (\ref{double}) shown in
the right panel of Figure \ref{fig:oneDEU}, we check that the method becomes
more accurate for intermediate and strong repulsion by increasing the size of
the cluster (see right panel inset). The deviation at weak coupling can in fact
be understood by a careful scrutiny of the method and its strengths. The
kinetic and interacting terms of the Hubbard Hamiltonian (\ref{Hub_Ham}) are
one- and two-body terms, respectively, when written in terms of the physical
fermions. In the mapped Hamiltonian (\ref{CBF_Ham}) part of the kinetic term is
computed in mean-field (inter-cluster contribution) and part is computed
exactly (intra-cluster contribution), while the on-site interaction is always
computed exactly.  As a result, when minimizing the effective mean-field
energy, the CFB mean-field tends to underestimate the kinetic contribution,
leading to a wave function with low double-occupancy at the expense of a higher
kinetic energy. By increasing the cluster size, we are including more hopping
processes into the exact computation. In this way, the method is able to yield
lower total energy by lowering the kinetic energy and adjusting the
double-occupancy to higher values. This trend is exactly analogous to
mean-field theory of physical fermions, where the HF method lowers the
one-particle kinetic energy to a large extent but yields too high a repulsion
energy and thus non-optimal total energy. In this way, it can be seen that the
CFB mean-field theory is highly complementary to the mean-field theory based on
physical fermions because they are treating different terms in a mean-field
way. The physical fermion mean-field theory performs better in the small $U$
regime, like in the Fermi-liquid phase, while the fermion-boson mean-field
theory performs better in large $U$ regimes like the  Mott insulating phase.
Also apparent in this figure is the systematic improvement of the method when
the size of the cluster is increased.

\begin{figure}[htbp]
\begin{center}
\includegraphics[width=0.49\textwidth]{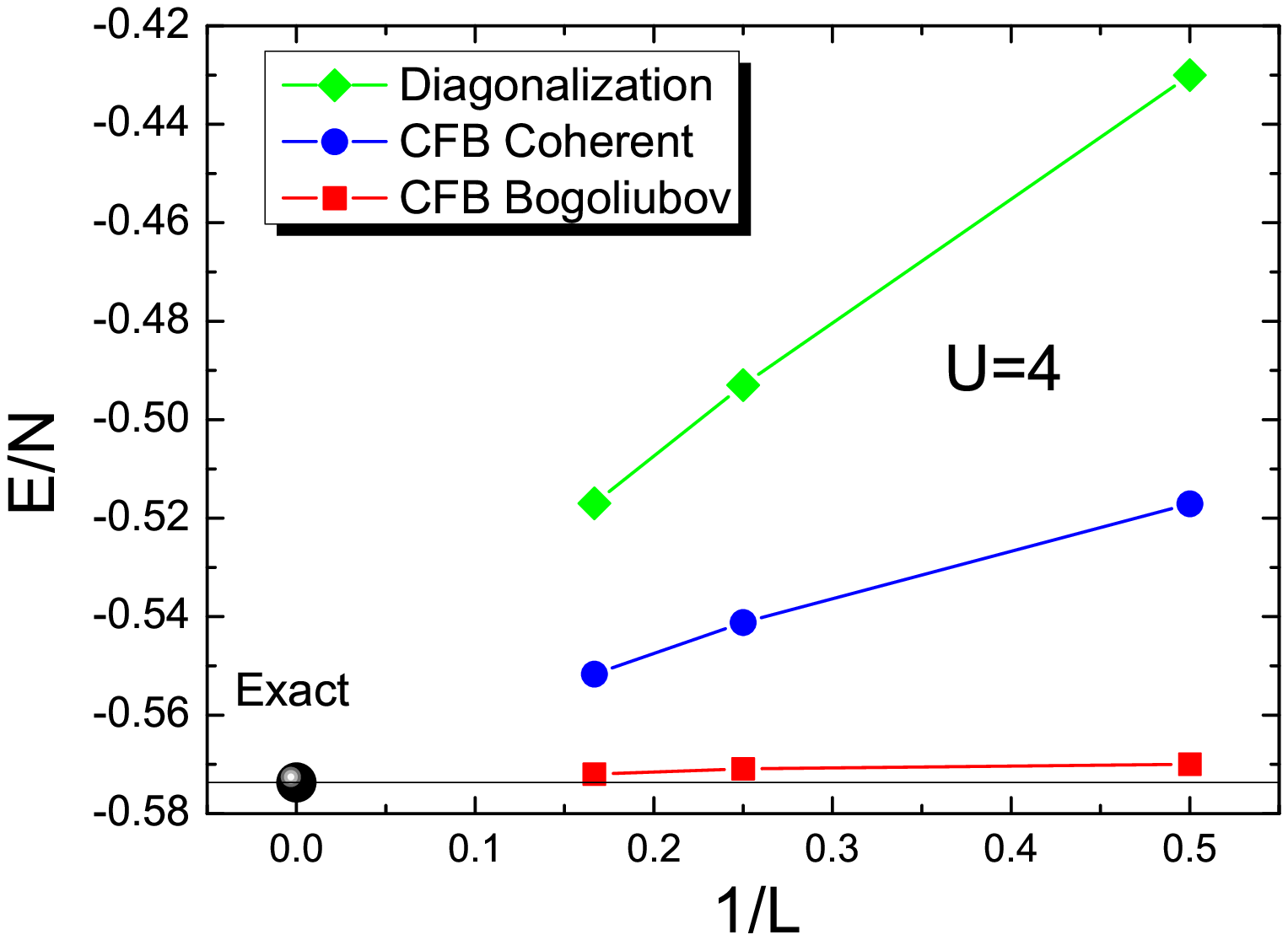}
\includegraphics[width=0.49\textwidth]{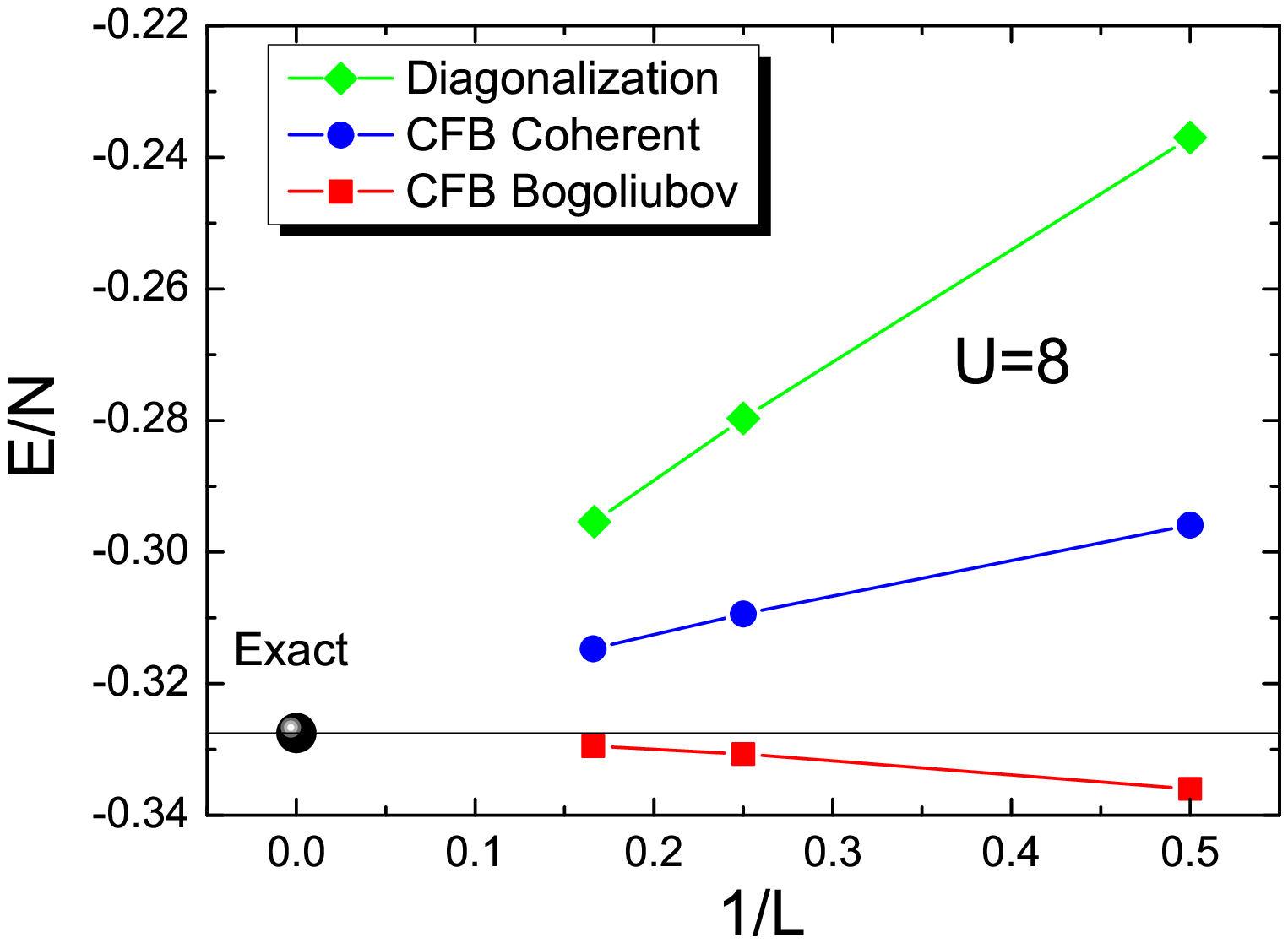}
\end{center}
\caption{Ground state energy per site versus the inverse of the clusters size
$L$ for the CFB approximations and exact diagonalization showing the convergence
to the exact Bethe ansatz value at intermediate ($U=4$) and strong ($U=8$)
Hubbard interactions.}
\label{figU4}
\end{figure}

In Figure \ref{figU4} we show the convergence of the energy per site towards
the exact Bethe ansatz result as a function of the reciprocal of the number of
sites in the cluster, $1/L$, for $U=4$ in the left panel and $U=8$ in the right
panel.  Specifically we depict the energy of a single isolated cluster obtained
from exact diagonalization of $\hat{H}_{R}^{\square}$, together with the two
CFB mean-field treatments described above: the bosonic coherent wave function
(CFB Coherent) and the bosonic Bogoliubov wave function (CFB Bogoliubov). A
clear improvement of the energy is seen in both panels as a function of $1/L$.
In the limit where the cluster becomes the entire system, inter-cluster
processes, which are here treated approximately in mean-field, are no longer
present, the CFB Hamiltonian becomes doubly quadratic and represents the even
and odd number parity sectors of the exact Hamiltonian. As a result, CFB
mean-field becomes exact in this limit and simply corresponds to picking the
exact ground state as the composite particle to occupy. As seen in Figure
\ref{figU4}, in both approaches (Coherent and Bogoliubov), the energy
extrapolates to the exact result. The  CFB Bogoliubov treatment adds bosonic
fluctuations to the coherent-state approximation. The improvement is apparent
in the figure, despite the fact that for strong coupling ($U=8$) the method
overbinds. This is due to violations of the physical constraint
(\ref{physical_constraint}) when treated on average (\ref{physical-average}),
that induces mixtures with unphysical states, a point that will be addressed in
future works. Notice that the computational cost associated with increasing the
cluster size is combinatorial, limiting the size of clusters that can be used
as building blocks for the mapping. An alternative way to improve the precision
of the method is the use of more sophisticated many-body approaches to treat
the CFB Hamiltonian (\ref{CBF_Ham}).

\begin{figure}[htbp]
\begin{center}
\includegraphics[width=0.49\textwidth]{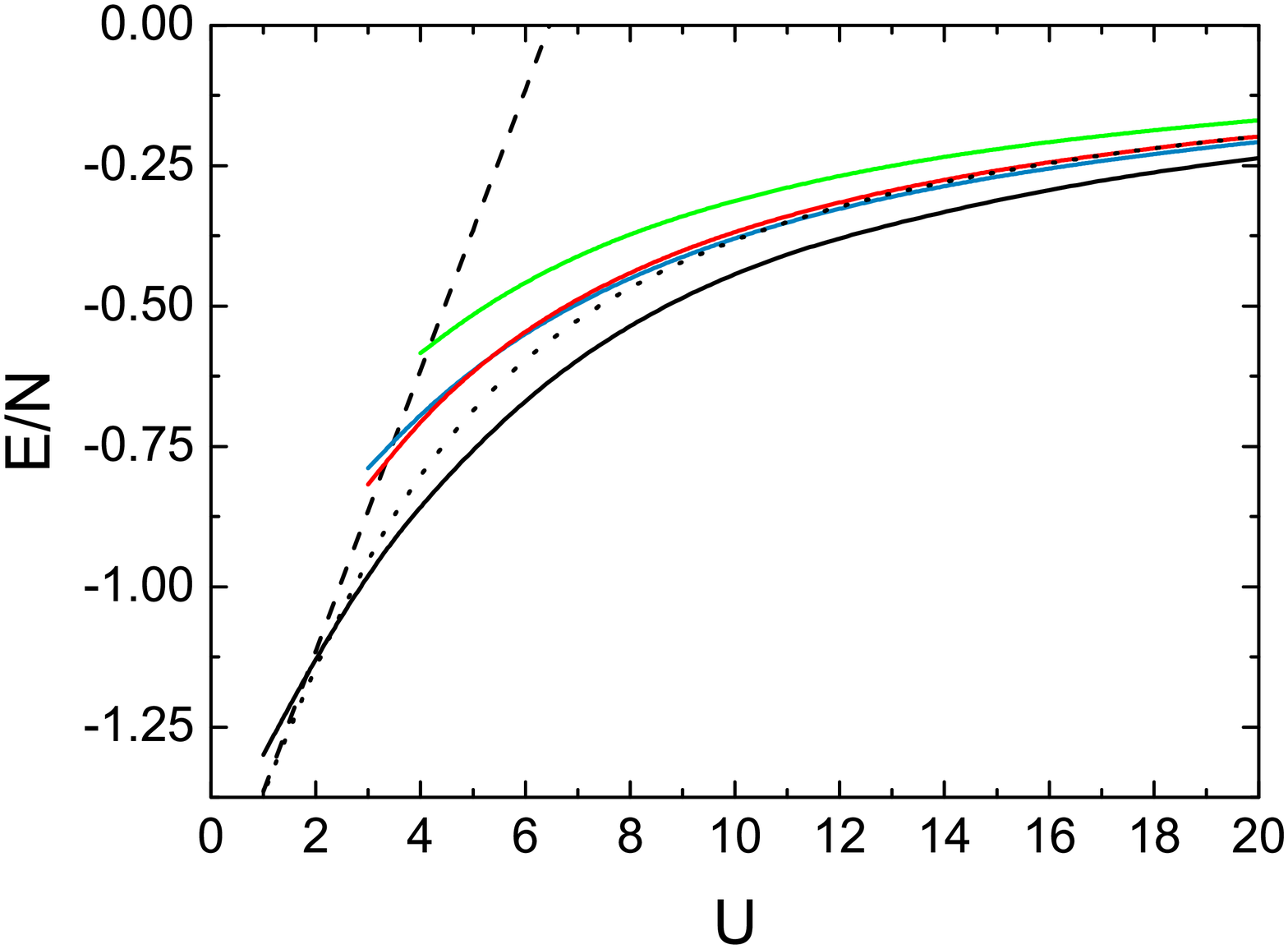}
\includegraphics[width=0.49\textwidth]{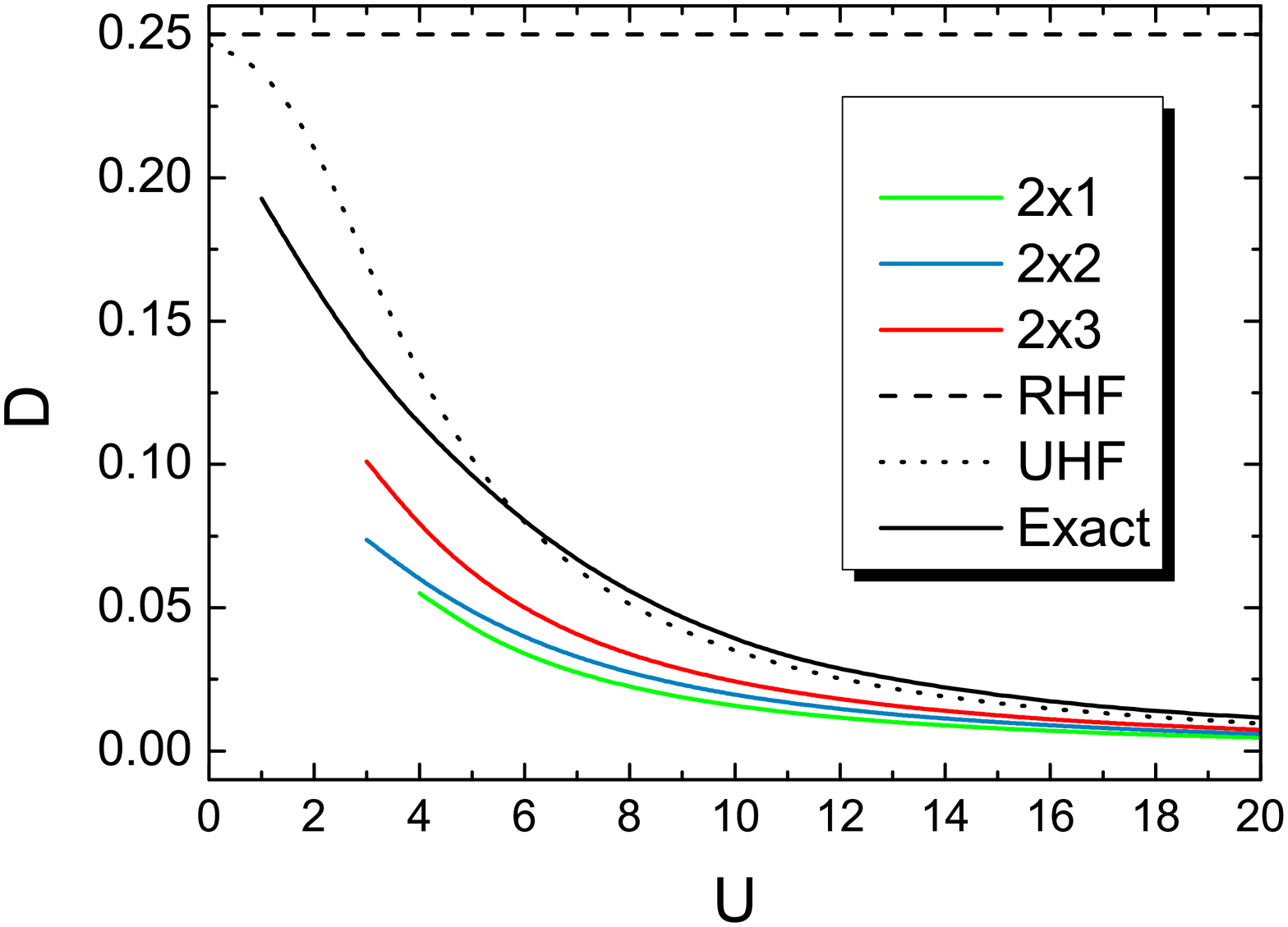}
\end{center}
\caption{
Ground state energies per site and double-occupancies of the 2-D Hubbard model
obtained from the CFB mean-field theory with various clusters sizes, from the
restricted and unrestricted Hartree-Fock approximation, and from an exact
diagonalization of an isolated $4 \times 4$ cluster. 
}
\label{fig:twoDEU}
\end{figure}

We have also applied the CFB mean-field approach within the coherent
approximation to the two-dimensional Hubbard model at half filling.  Results
for the ground state energies and double-occupancies are displayed in figure
\ref{fig:twoDEU} using clusters of sizes $2\times1$, $2\times2$, and
$2\times3$.  We have been unable to reach a self-consistent solution for small
$U$ values, even with sophisticated convergence-acceleration techniques
\cite{DIIS}. This might stem from the fact that larger spatial dimensionality
implies larger weight of the kinetic energy term in the Hamiltonian. In this
sense, when a larger part of the Hamiltonian is treated in mean-field,
convergence can be harder to achieve. The two-dimensional Hubbard model
exhibits antiferromagnetic (AF) long-range order for all repulsive $U$
\cite{AFMLO}. Inspecting our converged density matrix for the $2\times2$
cluster, we find that one of the two possible AF boson states gets a
significantly larger weight than the other. This symmetry breaking cannot be
observed in the $2\times1$ and $2\times3$ cases because these lattices are not
commensurate with AF long-range ordering.  This observation rationalizes the
significant energy difference between them and the reported relatively high
stability of the $2\times2$ case. Also the symmetry-breaking has not been
observed for the one-dimensional model, as expected.

A noticeable difference between the one-dimensional and two-dimensional cases is
the relative energy between the CFB mean-field theory and HF of physical
fermions. In 1D systems, starting from $U = 1.4$ for the size-six cluster,
$U=1.7$ for size-four cluster, and $U = 2.2$ for size-two cluster, HF yields a
lower ground-state energy than the composite mean-field approach. We can take
this $U$ crossover ratios as a rough estimate of the Mott transition point,
which the exact solution predicts at $U=0$ \cite{LiebAndWu}.  In the
two-dimensional problem, recent DMFT results \cite{Sordi} place the
Mott-insulator transition at approximately $U = 5.3$.  This is slightly higher
than the crossing points between HF and our composite mean-field results.

\section{Conclusion}

In this work, we have proposed a mapping of fermion operators onto composite
fermion and boson cluster operators in one-to-one correspondence to
many-electron states.  The mapping was used to transform the standard Hubbard
model onto a new composite fermion-boson Hamiltonian acting on a cluster
superlattice that can be treated by standard many-body approximations. The
advantage of the cluster mapping is that the short-range interactions and the
local quantum fluctuations are taken into account exactly from the onset.
Specifically, on-site interactions and intra-cluster hopping processes are
computed exactly by definition.

We have proposed a mean-field ansatz built as a tensor product of a fermionic
and bosonic wave function. This ansatz in terms of composite particles can
exactly handle single-cluster operators (such as the on-site repulsion), while
inter-cluster scattering processes (such as hopping) are treated in a mean-field
way.  We showed that the ansatz performs well for large $U$ at half-filling, and
is fairly complementary to the Hartree--Fock approach of physical fermions,
which treats hopping exactly at the expense of only a mean-field treatment of
the on-site repulsion interaction. We plan to explore, in future work, the
applicability of the ansatz to the doped phases of the Hubbard Hamiltonian.

\section{Acknowledgments}

This work was supported by the Department of Energy, Office of Basic Energy
Sciences, Grant No. DE-FG02-09ER16053, the Welch Foundation (C-0036), DOE-CMCSN
(DESC0006650), and by the Spanish Ministry of Economy and Competitiveness trough
grants FIS2012-34479 and BES-2010-031607.

\section*{References}
\bibliographystyle{unsrt}
\bibliography{FB}

\end{document}